\documentstyle[twoside,fleqn,espcrc2,epsf]{article}

 \newcommand{\pslash}{\not{\hspace{-0.08cm}p}}
\def\lsim{\mathrel{\rlap{\lower4pt\hbox{\hskip1pt$\sim$}}
    \raise1pt\hbox{$<$}}}                
\def\gsim{\mathrel{\rlap{\lower4pt\hbox{\hskip1pt$\sim$}}
    \raise1pt\hbox{$>$}}}                
\newcommand{\AmS}{{\protect\the\textfont2
  A\kern-.1667em\lower.5ex\hbox{M}\kern-.125emS}}

\title{\hspace{-0.25cm} Non-perturbative improvement and
 renormalization of lattice operators \hspace{-0.3cm} 
 \thanks{Talk given at Lattice '97 by P. Rakow.}}

\author{S.~Capitani%
           \address{Deutsches Elektronen-Synchrotron DESY,
                    D-22603 Hamburg, Germany},
        M.~G\"ockeler%
           \address{Institut f\"ur Theoretische Physik, Universit\"at
                    Regensburg, D-93040 Regensburg, Germany},
        R.~Horsley%
           \address{Institut f\"ur Physik, Humboldt-Universit\"at zu Berlin,
                    D-10115 Berlin, Germany},
        H.~Oelrich%
           \address{DESY-IfH Zeuthen, D-15735 Zeuthen, Germany},
        H.~Perlt%
           \address{Institut f\"ur Theoretische Physik, Universit\"at
                    Leipzig, D-04109 Leipzig, Germany},
        D. Pleiter$^{\rm d}$, 
        P.~E.~L. Rakow$^{\rm d}$,
        G.~Schierholz$^{\rm a,d}$
        A.~Schiller$^{\rm e}$ 
        and 
        P.~Stephenson$^{\rm d}$
}
       
\begin{document}

\begin{abstract}

The Alpha Collaboration has proposed an optimal
 value for $c_{SW}$ in the Sheikholeslami-Wohlert
action, chosen to remove $O(a)$ effects. To measure
 hadronic matrix elements to the same accuracy we need
a method of finding $O(a)$ improved operators, and
 their renormalization constants. We determine the $Z$ 
factors by a non-perturbative method, measuring the
 matrix elements for single quark states propagating
through gauge fields in the Landau gauge. The data show
 large effects coming from chiral symmetry breaking.
This allows us to find the improvement coefficients too, 
 by requiring that the amount of chiral symmetry
 breaking agrees with that predicted by the chiral Ward identities. 

\end{abstract}

\maketitle

\section{INTRODUCTION}

   There is current interest in improving the Wilson fermion action
 to remove all $O(a)$ discretization errors. An improved action is
 already known \cite{alpha_csw}, however to measure matrix elements
 without $O(a)$ errors we also need a method of
 improving the operators. This is done by adding improvement 
 terms, operators of higher dimension, either with one extra 
 derivative or with an extra power of the quark mass $m_q$. 
   In this talk we discuss a non-perturbative method of determining 
 the renormalization factors, $Z$, and the improvement coefficients, 
 $c_i$ and $b$, for these operators. 

 The method of finding $Z$ is that suggested in
 \cite{Martinelli}, operator expectation values are measured for 
 quark states travelling through Landau gauge gluon configurations.
 The $Z$ factors
 are found from the ratio of the measured expectation values to the 
 tree-level value. The improvement coefficients can be found from
 the same sort of data, by using the chiral Ward identities 
 of QCD. 
 Our computational methods are discussed in more detail
 in \cite{OelrichL97}.
 \vspace*{-21cm} \\ DESY 97-180 \\ HUB-EP-97/67 \vspace*{20cm} 

\section{CHIRAL WARD IDENTITIES} 

   We need a way to identify the $O(a)$ effects in three-point 
 functions, so that we can tune the improvement parameters to 
 remove them. There is a way to do this, since the Wilson mass term
 and the Sheikholeslami-Wohlert clover term both violate chiral 
 symmetry. 

\begin{figure}[tbh]
\vspace*{-0.3cm}
\epsfxsize=6.5cm \epsfbox{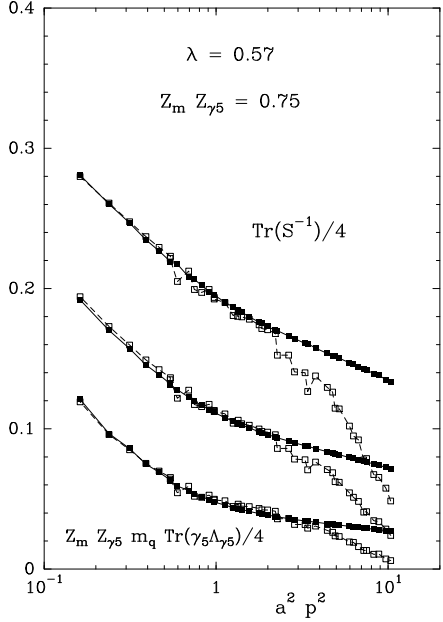}
\vspace*{-0.8cm}
\caption{\footnotesize The chiral Ward identity
 Eq.(\ref{clovpropward}) relating
 the fermion propagator (open symbols) to the pseudoscalar density
 (filled symbols). The data are for $\beta = 6.0, c_{SW} = 1.769$,
 and the $\kappa$ values $0.130, 0.1324$ and $0.1342$.}
\vspace*{-0.5cm}
\label{fig:fermprop}
\end{figure}

    If we were simulating a theory which respected chiral symmetry,
 it would be easy to use this fact to pick out $O(a)$ effects, and 
 we could recognise the correct improvement parameters by the restoration 
 of chiral symmetry. However for lattice QCD things are not quite this
 simple. As well as the chiral symmetry violation coming from lattice 
 artifacts, which we want to remove, there is also genuine chiral
 symmetry violation, some coming from the fact that we work with 
 finite quark mass, and some from the well-known fact that QCD  
 spontaneously breaks chiral symmetry. We need to know how big the
 real violation of chiral symmetry is. 

    Fortunately there is a way to identify this genuine component
 of chiral symmetry breaking \cite{Sharpe97}. Chiral Ward
 identities can be 
 constructed for any Green function, which tell us exactly how
 much violation of chiral symmetry should be present. 
 
  The simplest case to consider is the fermion propagator. 
 In theories, such as the naive fermion action,  where the
 only explicit breaking of chiral symmetry
 is a mass term $m_0 \overline{\psi} \psi$, 
 the quark propagator satisfies \cite{Pagels}
 \begin{eqnarray}
 \gamma_5 S(p) + S(p) \gamma_5 = 2 m_0 G_{\gamma_5}(p) 
  \nonumber
 \\ \Rightarrow 
 \gamma_5 S^{-1}(p) + S^{-1}(p) \gamma_5 = 2 m_0 \Lambda_{\gamma_5}(p),  
 \label{propward} 
 \end{eqnarray}  
 where $\Lambda_{\gamma_5}$ is the amputated three-point function.
 Because $\pslash$ and $\gamma_5$ anticommute, this identity  
 relates the running mass (the scalar part of $S^{-1}$) to the
 pseudoscalar density.  
 With naive fermions $ Z_m Z_{\gamma_5} = 1$, so the Ward identity
 Eq.(\ref{propward}) holds in the same form  
 both for bare and renormalized Green functions.
 If chiral symmetry is spontaneously broken the right hand side
 doesn't vanish as $m_0 \to 0$, and the fermion propagator looks
 massive even in the chiral limit \cite{Pagels}.    

   In our action this chiral Ward identity is not automatically true.    
 As well as chiral symmetry breaking from the mass term there is 
 additional breaking from the Wilson and clover terms.
  We can now determine improvement 
 coefficients, and find some information about the $Z$s, 
 if we impose the chiral Ward
 identity, by insisting that the improved 
 Green functions obey the identity 
 \begin{equation}
 \gamma_5 S^{-1}_{imp}(p) + S^{-1}_{imp}(p) \gamma_5 = 
  Z_m Z_{\gamma_5} 2 m_q \Lambda^{imp}_{\gamma_5}(p), 
 \label{clovpropward} 
 \end{equation} 
 where $m_q \equiv 1/(2 \kappa) - 1/(2 \kappa_c)$.

   We are fortunate that improving $G_{\gamma_5}$ is easier than    
 improving most operators. There is no single-derivative operator
 to add, all we 
 have to do is multiply  $G_{\gamma_5}$ by a factor of the type
 $(1 + b a m_q)$. 

   The improvement of the quark propagator is slightly more
 complicated. As well as multiplying by a $m_q$-dependent factor, 
 we also have to subtract a contact term from the
 propagator ~\cite{Martinelli,Heatlie,CapitaniL97}
\begin{eqnarray*}
 S(p)  = \frac{Z_2}{( 1 + b_2 a m_q)}
 \frac{1}{(i \pslash + m_R(p))} +  a  \lambda 
   + O(a^2)&& 
 \vspace*{-0.3cm} 
\end{eqnarray*} 
\begin{eqnarray}
&\Rightarrow& 
 S^{imp}(p)  = ( 1 + b_2 a m_q ) \left( S(p) - a \lambda \right).
\label{Simprove}
\end{eqnarray} 
 $\lambda$ is a constant in momentum space, or a $\delta$ function 
 in position space. This makes the chiral symmetry breaking term as 
 short-range as possible. In perturbation theory we can see that the $O(a)$
 corrections to the propagator only take this simple form if $c_{SW}$
 is correctly chosen ~\cite{CapitaniL97}. 

   In Fig.\ref{fig:fermprop} we show how well the Ward identities
 can be satisfied when the improvement coefficients are correctly
 chosen. The agreement is good while $a^2 p^2 \lsim 2$. 

\section{THE LOCAL VECTOR CURRENT}

  There are similar Ward identities for the flavour non-singlet 
 three-point functions, but
 they require the measurement of four-point functions $H$.
\begin{eqnarray}
 H_{\Gamma;\gamma_5}(p) \equiv 
 \sum_{ijkl} \langle M^{-1}_{ij} \Gamma  
  M^{-1}_{jk} \gamma_5  
 M^{-1}_{kl} \rangle e^{i p (x_i - x_l)}
 \nonumber \\
 H_{\gamma_5;\Gamma}(p) \equiv \sum_{ijkl} 
 \langle M^{-1}_{ij} \gamma_5 
  M^{-1}_{jk} \Gamma  
 M^{-1}_{kl} \rangle e^{i p (x_i - x_l)}
 \label{Hdef}
\end{eqnarray}
 where $M$ is the fermion matrix. In naive fermions 
  Ward identities analogous to Eq.(\ref{propward})
 relate the chiral violation in the 
 three-point functions $G$ to $m_0 H$. For example, for the local 
 vector current,  
 \begin{equation}
   \gamma_5 G_{\gamma_\mu} + G_{\gamma_\mu} \gamma_5 = 
 m_0 \left( H_{\gamma_\mu;\gamma_5} + H_{\gamma_5;\gamma_\mu} \right).
 \label{eq:wardGH}
 \end{equation} 
 The $H$s also appear in Ward identities for the parity-splitting 
 between the operators $\Gamma$ and $\Gamma \gamma_5$.

\begin{figure}[tbh]
\vspace*{-0.3cm} \hspace*{0.3cm}
\epsfxsize=6.5cm \epsfbox{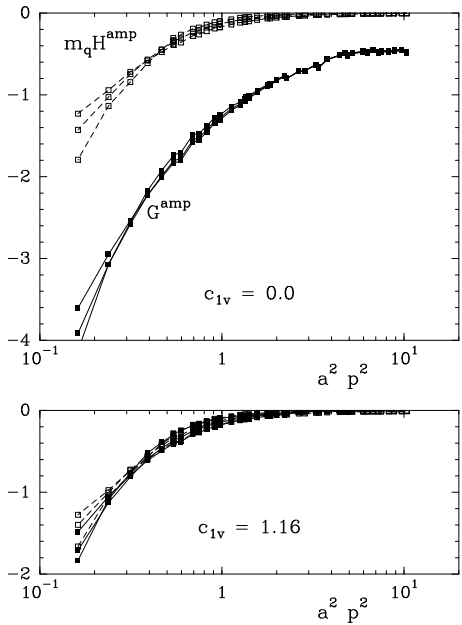}
\vspace*{-0.8cm} \hspace*{-0.3cm} 
\caption{\footnotesize The left and right hand sides of the 
 chiral Ward identity Eq.(\ref{eq:wardGH}) for the improved
 vector current, with $c_{1v} = 0$ (upper figure) and $c_{1v} = 1.16$ 
 (lower figure). When we choose $c_{1v}$ correctly, the Ward identity
 holds well.  The $\kappa$ values are the same as Fig.1.}
\vspace*{-0.5cm}
\label{fig:vector}
\end{figure}

 Again these identities do not hold automatically for clover fermions,
 but we can require that they hold for the improved and renormalized 
 $G$s and $H$s. We should point out that the renormalization of
 the $H$s is not simply multiplicative. 
 Because the sum over the sites in Eq.(\ref{Hdef})
 includes short-distance contributions, coming from regions where
 the two operator-insertions are close together,
 additional counter-terms, proportional to three-point functions, 
 arise. These will be discussed in more detail in ~\cite{future}. 
 
    The first operator we have tried this method on is the local 
 vector current. In Fig.~\ref{fig:vector} we compare the  
 amputated Green function  for the improved local vector current, 
 with the value it should have, according 
 to the Ward identity Eq.(\ref{eq:wardGH}). As in \cite{Martinelli}
 amputations are performed by dividing by the improved fermion
 propagator. 
 We see that the comparison is very sensitive 
 to the value of the improvement coefficient $c_{1v}$, defined by 
 \begin{equation}
 \left( \overline{\psi} \gamma_\mu \psi \right)^{\rm{imp}} \! \! =
 (1 + a b m_q)  \overline{\psi} \gamma_\mu \psi -
 a c_{1v} \overline {\psi} \stackrel {\leftrightarrow}{D_\mu} \psi. 
 \end{equation}  
 When we use a poor value (as in the upper graph) the violation of 
 Eq.(\ref{eq:wardGH}) is clear. 
 Our preliminary result is that we find the best agreement with the 
 Ward identity when $c_{1v} \approx 1.16, b \approx 0.1$ and
 $Z_v \approx 0.87$. Since we are working with off-shell 
 quantities we can not use the equations of motion to eliminate
 either $b$ or $c_{1v}$. 

 We can  
 test these coefficients, derived from off-shell quarks, by using 
 them to renormalize our on-shell nucleon measurements. The results 
 are in good agreement with the known value of the conserved
 vector current \cite{HorsleyL97}. 
 So we see that using chiral Ward identities to tell us when $O(a)$ 
 effects have been removed seems a promising method. 

\section*{ACKNOWLEDGEMENTS}

This work was supported in part by the Deutsche Forschungsgemeinschaft. The
numerical calculations were performed on the Quadrics computers at 
DESY-IfH and Bielefeld University.

\end{document}